\documentstyle[aps,twocolumn,epsf]{revtex}

\begin{document}

\title{New Class of Random Matrix Ensembles with Multifractal
Eigenvectors.} 
\date{\today}
\author{ V.\ E.\ Kravtsov$^{1,2}$ and K.\  A.\  Muttalib$^{3}$.}
\address{$^1$ International Center for Theoretical Physics, P.O. Box
586, 34100 Trieste, Italy\\$^2$
Landau Institute for Theoretical Physics, Kosygina str. 2, 117940 Moscow,
Russia\\ $^3$ Department of Physics, University of Florida, Gainesville,
Florida 32611}
\maketitle
\begin{abstract}
Three recently suggested random matrix ensembles (RME) are linked together
to represent a class of RME with multifractal eigenfunction statistics.
The generic form of the two-level correlation function for the case of
weak and extremely strong multifractality is suggested.
\end{abstract}
\draft
Random matrix ensembles turn out to be a natural and convenient
language 
to formulate generic statistical properties of energy levels and
transmission matrix elements in complex
quantum systems. Gaussian random matrix ensembles, first introduced by
Wigner and Dyson $^{1,2}$ for describing
the spectrum of complex nuclei, became very popular in solid state
physics as one of the main theoretical tools to study mesoscopic
fluctuations $^{3}$ in small disordered electronic systems. The success of
the random
matrix theory (RMT) $^{2}$ in mesoscopic physics is due to its extension
to
the problem of electronic transport based on the Landauer-B\"uttiker
formula $^{4}$ and the statistical theory of
transmission eigenvalues $^{5}$. Another field where the RMT is exploited
very intensively is the problem of the semiclassical approximation in
quantum systems whose classical counterpart is chaotic $^{6}$. It turns out
$^{6}$
that the energy level statistics in true chaotic systems is described by
the RMT, in contrast to that in the integrable systems where in most
cases it is close to the Poisson statistics.

Apparently the nature of the energy level statistics is related to the
structure of eigenfunctions, and more precisely, to the overlapping 
between different eigenfunctions. This is well illustrated
by spectral statistics in a system of
non-interacting electrons in a random potential which exhibits the
Anderson
metal-insulator transition with increasing disorder.
At small
disorder the
electron wave functions are extended and essentially structureless.
They overlap very well with each other resulting in energy level repulsion
characteristic of the Wigner-Dyson statistics. On the other hand in the
localized phase electrons are typically localized at different points of
the sample, and in the thermodynamic limit where the system size
$L\rightarrow\infty$ they ``do not talk to each other''. In this case there
is no correlation between eigenvalues, and the energy levels follow
the Poisson statistics. 

The energy level statistics in the critical region near the
Anderson transition turns out to be universal and different from both
Wigner-Dyson
statistics and the Poisson statistics $^{7,8}$. 
A remarkable feature of the critical level statistics
is that the level number variance $\Sigma_{2}(\bar{N})=\langle
(\delta N)^{2}\rangle=\chi \bar{N} $ is
asymptotically
linear in the mean number of levels $\bar{N}\gg 1$ in the energy window.
Such a quasi-Poisson behavior was first predicted in Ref.[9]. Later
the existence of the linear term in $\Sigma_{2}(\bar{N})$  was
questioned $^{8}$, since for this term to appear the normalization sum rule
should be violated. 
It has been shown recently $^{10}$ that the new qualitative feature 
responsible for
the violation of the sum
rule and the existence of the finite ``level compressibility'' $\chi$ is the
{\it multifractality} of critical wave functions $^{11,12}$. 

The notion of multifractality 
is two-fold. The first (and most widely accepted) property of
multifractality is related with the space structure of a {\it single}
wave function $\Psi_{n}({\bf r})$. It is defined through the moments of
inverse participation ratio $^{11}$:
\begin{equation}
\label{IPR}
I_{p}=\sum_{{\bf r}}\langle |\Psi_{n}({\bf r})|^{2p} \rangle\propto
L^{-D_{p}(p-1)},   
\end{equation}
where $L$ is the system size, $d$ is the dimensionality of space;
 $p>1$ is an integer.
The set of exponents $D_{p}< d$
characterize   
the fractal dimensionality of the cluster where $|\Psi_{n}({\bf r})|$ is
larger
than a certain value that increases with increasing $p$.

The second, far less appreciated property of multifractality, is related
to the overlapping of ${\it different}$ wave functions with energies
$E_{n}$ and $E_{m}$. 
The main effect
of multifractality on spectral statistics is given by the simple
overlapping of local densities
($p=2$).
The corresponding fractal dimensionality
$D_{2}$ is the most
important critical exponent.
For $|E_{n}-E_{m}|\gg \Delta$
($\Delta=1/\bar{\rho} L^d$, where $\bar{\rho}=\langle 
\rho(E)\rangle$ is the 
mean density of states) the form of the local
density correlation function
has been suggested and confirmed numerically in Ref.[12]:
\begin{equation}
\label{ldc}
\langle |\Psi_{n}({\bf r})|^{2}  |\Psi_{m}({\bf r})|^{2}
\rangle
\propto |E_{n}-E_{m}|^{-(1-\frac{D_{2}}{d})}.
\end{equation}
A remarkable feature of multifractality is that the local density
correlation function decreases very slowly with increasing $|E_{n}-E_{m}|$
so that two fractal wave functions, however sparse they are, should still
overlap strongly $^{13}$. 

One of the consequences $^{10}$ of Eq.(\ref{ldc}) is the anomalous
Poisson-like
term in the level number variance $\Sigma_{2}(\bar{N})$ which is
characterized by the level compressibility $\chi$:
\begin{equation}
\label{chi}
\chi=\frac{d-D_{2}}{2d}.
\end{equation}
It is immediately seen from Eq.(\ref{chi}) that the critical level
compressibility
never reaches the Poisson limit $\chi=1$. For an infinitely sparse
fractal $d-D_{2}\rightarrow d$ is maximal, yet $\chi$ is equal to 1/2 and
not to 1.
This is because even the infinitely sparse critical fractal eigenfunctions
overlap strongly,
in contrast to two localized states $^{10,13}$.

One may assume that the universal critical level statistics which is
described by a set of critical exponents $D_{p}$, applies to a wider class
of physical systems and it is in fact generic for an intermediate
situation between chaos and integrability. An example of such a system
has been proposed recently $^{14}$. It turns out that the 
Coulomb impurity
inside an integrable square billiard leads to a
drastic reconstruction of eigenstates however small is the strength
of the potential.  In such a "Coulomb billiard" all eigenfunctions in the
momentum representation exhibit multifractality. 

It is therefore natural to look for a RME with
multifractal
eigenvector and eigenvalue statistics similar to that at the
mobility edge in disordered electronic systems.
Such a RMT would provide a 
description of generic features of the critical level statistics. 

One such ensemble is suggested in Ref.[15]. It is the Gaussian ensemble of
$M\times M$ hermitian matrices $H$ with independent random entries ($i\geq
j$) defined
by:
\begin{equation}
\label{I}
\langle H_{ij}\rangle=0,\;\;\;\; \langle(H_{ij}^{\mu})^{2} \rangle = 
\beta^{-1}[1+ 
|i-j|^{2}/B^2 ]^{-1},
\end{equation} 
where $H_{ij}^{\mu}$
are real random
components ($\mu=1$ for $i=j$, $\mu=1,...\beta$ for $i> j$);
$\beta=1,2,4$ for Dyson's orthogonal, unitary and symplectic
ensembles, and $B$ is a parameter. For $B\gg 1$ this RME can be mapped
onto a nonlinear supersymmetric sigma-model $^{15}$. The case $B\ll 1$
corresponds to 2D Coulomb billiard considered in Ref.[14].
The presence of
multifractality, Eq.(\ref{IPR}), and Eq.(\ref{chi}) has
been proved $^{14,15}$ for this RME. 
In what follows we will use this RME (RME-I) as a reference point.

There are two more RMEs $^{16,17}$ which 
were suggested recently as possible candidates to describe
the critical level statistics. However, their definitions  are
so drastically different 
that they were considered as two {\it alternative}
options albeit the two-level correlation
functions (TLCF)
$R(\varepsilon,s)=\langle\rho(\varepsilon)\rho(\varepsilon+s)\rangle_{c}$
in the proper regimes are {\it identical} for both
RMEs. It was first noted in Ref.[18] that
since the energy level statistics is a "fingerprint" of the statistics of
eigenfunctions, the latter in the corresponding regimes of these two
models 
should also be similar.

The first quantitative link between the
predictions of RME equivalent to that studied in Ref.[16] (RME-II) and 
numerics on the 3D Anderson model at the mobility
edge has been done in Ref.[19]. Surprisingly enough it was possible 
to fit very well
the numerics for the critical level spacing distribution $P(s)$
in 3D Anderson model by a proper choice of only one parameter in RME-II.
Moreover, the level compressibility $\chi$ in the RME-II for this
particular choice of parameter turned out to be very close to that found
numerically for the 3D Anderson model. 

In this Letter we argue that  RME-II and a certain
critical regime in RME-III
studied in Ref.[17] are equivalent to RME-I and thus possess the
multifractality. Altogether they form a new class of RME which
describes certain remarkable features of critical level statistics.

We start with the definitions of RMEs studied in Ref.[16,17]. The
probability density $P(H)$ for a $M\times M$ random
Hamiltonian $H$ 
from RME-II is given by:
\begin{equation}
\label{II}
P_{{\bf II}}(H)\propto \exp[-\beta Tr V(H)],
\end{equation}
where
the ``confinement potential'' $V(H)$
grows extremely slowly with $H$:
\begin{equation}
\label{ab}
V(H)=\frac{1}{\gamma}\,\ln^{2} H,\;\;\;\;H\gg 1.
\end{equation}
This is crucial for the universality of the eigenvalue
statistics in the limit  
$M\rightarrow\infty$ $^{20}$. For $V(H)$ growing
slower
than $H$ the full universality is no longer present, $^{21,22}$ and the
eigenvalue statistics may, and does differ from the Wigner-Dyson statistics 
$^{16}$.
Another important feature of Eq.(\ref{II}) is that the distribution
function $P_{{\bf II}}(H)$ is invariant under the rotation of basis
(unitary invariance):
\begin{equation}
\label{ut}
P_{{\bf II}}(H)=P_{{\bf II}}(UHU^{\dagger}).
\end{equation}

In contrast to Eq.(\ref{II}), the distribution function for RME studied in
Ref.[17] is 
Gaussian. However, the unitary invariance  is broken
by a fixed unitary matrix $\Omega$:
\begin{equation}
\label{III}
P_{{\bf III}}(H)\propto e^{-\frac{\beta}{2} Tr
H^2}\,e^{-\frac{\beta}{2}\,b\,Tr\{[\Omega,
H][\Omega,H]^{\dagger} \}}.
\end{equation} 
The properties of this RME depend on the choice of $\Omega$. For the
reasons discussed below we consider as RME-III the RME defined by
Eq.(\ref{III}) with $\Omega=diag(e^{i\theta_{j}})$, where
$\theta_{j}=2\pi j/M$.
The relevant critical regime for this RME corresponds to the
symmetry
breaking field $b$ that scales with 
$M\rightarrow\infty$
as $b=h^2 M^{2}$, where $h$ is a parameter.

Now it is clear why the RMEs given by Eqs.(\ref{II}),(\ref{III}) seem so
drastically different. The lack of unitary invariance of $P_{{\bf
III}}(H)$ means a
preferential
basis. The existence of such a basis implies a certain structure of 
eigenfunctions (in this basis) which should lead to spectral
statistics different from the Wigner-Dyson one. 
However, it seems
there is no way to get any structure of eigenfunctions in the {\it
unitary-invariant} RME-II. It follows immediately from Eq.(\ref{ut}) that
the distribution function $P_{{\bf II}}(H)$ depends only on
$E_{n}$, and the statistics of
eigenfunctions in RME-II is trivial and the same as for standard Gaussian
ensembles $^{2}$. Then the physical picture
that the spectral statistics is related to the statistics of
overlapping eigenfunctions seems to leave only one single option: the
Wigner-Dyson
energy level statistics in RME-II.

Nonetheless TLCF
$R(\varepsilon,s)=\delta(s)+Y_{2}(\varepsilon,s)$
proves to be {\it identical}
in these RMEs and
 after unfolding $^{23}$ it takes the form $^{16,17}$:
\begin{equation}
\label{TLCF}
Y_{2}(\varepsilon,s)\approx -
\frac{\pi^2 \eta^2}{4}\,\frac{\sin^{2} [\pi s]}{\sinh^{2}
[s\pi^2 \eta/2]},\;\;\;(\beta=2),
\end{equation} 
where   $\eta=\gamma/\pi^2 \ll 1$ or $\eta=h\ll 1$ for
RME-II and RME-III,
respectively, and  $\varepsilon\gg |s|$.
Eq.(\ref{TLCF}) coincides with the density correlation function
for a free electron gas at a {\it finite} temperature
$\eta\varepsilon_{F}$
and differs from the RMT result.

The way out from this contradiction is suggested in Ref.[18] where it was
conjectured that the unitary invariance is broken in RME-II {\it
spontaneously}. This means that the statistics of eigenfunctions in this
ensemble should be calculated {\it after} an infinitesimal
symmetry-breaking
term similar to that in Eq,(\ref{III}) is added. Then the identical 
TLCF in RME-II and RME-III should be considered
as an evidence that the proper procedure should result in
similar
eigenfunction statistics in RME-II and RME-III.

The progress $^{17}$ in studying the level statistics in RME-III that lead
to Eq.(\ref{TLCF}) is due
to averaging over the unitary group $\Omega$. The level statistics depend
on the configuration of eigenvalues $e^{i\theta_{j}}$ of $\Omega$.
The main contribution to
the average is made by $\Omega$ with the most homogeneous configurations
of $\theta_{i}$, the property known as an eigenvalue repulsion $^{2,17}$
Therefore, one may expect that the spectral statistics 
obtained {\it after} such an averaging is close to that corresponding to
a {\it single} unitary matrix $\Omega$ with eigenvalues
$\Omega_{j}=\exp[(2\pi i/M)j]$ (RME-III). As a matter of fact for $h\ll 1$
it turns out to be the {\it same}.

In order to show that we note that 
in the limit $M\rightarrow \infty$ Eq.(\ref{III}) leads to:
\begin{equation}
\label{corrs}
\langle (H_{ij}^{\mu})^{2}\rangle= \frac{1}{\beta}\,
\frac{1}{1+\frac{4\pi^2 b}{M^2}\,|i-j|^{2}}.
\end{equation}
If $b/M^2\rightarrow 0$, then we have a standard Gaussian ensemble
$^{1,2}$ and the Wigner-Dyson statistics. In the opposite case
$b/M^2\rightarrow \infty$, we have an ensemble of random diagonal matrices
and the Poisson statistics. In the critical case considered here for $b=h^2
M^2$, the  behavior of $\langle (H_{ij}^{\mu})^{2}\rangle$ is the same
as in Eq.(4) defining RME-I. We conclude that  the $M\rightarrow
\infty$ limits of RME-I and RME-III   
coincide. Then TLCF for RME-III and RME-I must be identical.

Fortunately, the latter can be calculated directly.
TLCF can be expressed $^{24}$ in terms of the spectral
determinant $P(s)$ as follows:
\begin{equation}
\label{R-P}
Y_{2}(s)= -\frac{1}{2\pi^2 s^2}- \frac{1}{4\pi^2}\,\frac{d^2}{ds^2}\ln 
P(s)+\frac{\cos(2\pi s)}{2\pi^2 s^2}\,P(s),
\end{equation} 
where
\begin{equation}
\label{P}
P(s)=\prod_{n\neq 0}\left(1+\frac{s^2}{\epsilon_{n}^2} \right)^{-1},
\end{equation}
and $\epsilon_{n}$ is a spectrum of the 
quasi-diffusion modes. The latter
can be found from the mapping $^{15}$ onto the nonlinear sigma-model
($B\gg 1$)
as follows: $\epsilon_{n}=4B |n|$, where for the periodic boundary
conditions $n=\pm 1, \pm 2, ...$. Making use of the identity $x^{-1}\sinh
x=
\prod_{n=1}^{\infty}(1+x^2/\pi^2 n^2)$ we immediately arrive at
Eq.(\ref{TLCF}) with $\eta=1/(2\pi B)$.   
Using the results of Ref.[15] one
can express the multifractality exponent $D_{2}$ in terms of $B\gg 1$.
For $\beta=2$ it appears to be $D_{2}=1-1/(2\pi B)$ which helps to
identify the
parameter $\eta$ in Eq.(\ref{TLCF}) as $\eta=1-D_{2}$. Thus all three
ensembles share the
same TLCF, Eq.(\ref{TLCF}) which is generic
to RME with weak multifractality $\eta\ll 1$.

The level
compressibility $\chi$ in Eq.(\ref{chi}) is obtained by the
integral
of the TLCF $^{8,10}$: 
\begin{equation}
\label{sr}
\chi=1+\int_{-\infty}^{+\infty}Y_{2}(\varepsilon,s)\,ds.
\end{equation}
Using Eq.(\ref{TLCF}) one can calculate 
the level compressibility $\chi=\eta/2+[1-\coth(2/\eta)]$
in the limit of weak multifractality $\eta\ll 1$. Neglecting the
exponentially small terms, we observe that Eq.(\ref{chi}) (with $d=1$)
is fulfilled. 

Note that both the linear level number variance $\Sigma_{2}(\bar{N})$
with $\chi\neq 1$ and TLCF of the form Eq.(\ref{TLCF}) are {\it not} the
trivial
consequences of the basis preference. A good counter-example is a Gaussian
RME with  the variance of the fluctuating diagonal components 
different from that of the off-diagonal ones. Their ratio $\mu=M^2
/\lambda^2$ 
sets the new energy scale $\lambda\gg 1$ in the problem, such that for
$s\gg
\lambda$ spectral correlations deviate from the Wigner-Dyson form.
However, the recent analytical results $^{25,26}$ show that this deviation
is
qualitatively different from that described by Eq.(\ref{TLCF}) for $s\gg 
1/\eta$. Albeit the oscillations in $Y_{2}(s)$ die out for $s\gg \lambda$,
there is still
left a constant tail $Y_{2}(s)=1/\pi^2 \lambda^2$ that extends up to
$s=\lambda^2$. 
Therefore the level number
variance $\Sigma_{2}(\bar{N})=(\bar{N}/\pi\lambda)^2$ for $N\ll
\lambda^2$ and $\Sigma_{2}(\bar{N})=\bar{N}$ for $N\gg \lambda^2$.

With increasing $\gamma$ and $h$ or decreasing $B$ the fractal
dimensionality $D_{2}$ decreases and Eq.(\ref{TLCF}) is no
longer valid. 
It is reasonable to assume that in the limit $h,\gamma\rightarrow\infty$
or
$B\rightarrow 0$ the fractal eigenvector becomes infinitely sparse, 
$D_{p}\rightarrow 0$. 
For RME-I this is, indeed, the case $^{14}$.
In this limit 
Eq.(\ref{chi})
predicts $\chi=1/2$. Let us check this prediction using an exact form of
TLCF given in Ref.[16]. 

First of all we note that even after unfolding, the TLCF
$R(\varepsilon,s)$ in RME-II is not
invariant under a shift in $\varepsilon$. 
In the limit $\gamma\rightarrow\infty$ the TLCF has
the same form in the
orthogonal, unitary, and symplectic ensembles $^{19,22}$:
\begin{equation}
\label{lim}
Y_{2}(\varepsilon,s)=-\theta(1/4-|s-\delta|), 
\end{equation}
where $-1<4\delta <1$ is a deviation of $4\varepsilon\gg 1$ from
the nearest odd-integer, and $\theta(x)$ is a step-function.

The lack of translational invariance is a peculiarity of the particular
RME-II. Only the TLCF {\it
smoothened} by averaging over $\delta$ can be physically meaningful. 
So we arrive at the TLCF of the
triangle form:
\begin{equation}
\label{tr}
Y_{2}(s)=\left\{\begin{array}{ll} 2|s|-1, & |s|<1/2,\\
0 & {\rm otherwise}.
\end{array} \right. \;\;\;(\beta=1,2,4).
\end{equation}
It is remarkable that after the substitution into Eq.(\ref{sr}) of either
Eq.(\ref{lim}) or Eq.(\ref{tr}) we get the predicted value $\chi=1/2$.
One may thus expect $^{27}$ that the triangle form of the TLCF is generic 
for all critical states in the limit of an infinitely sparse fractal 
$D_{2}\rightarrow 0$ $^{28}$. 

Let us discuss the applicability of Eqs.(\ref{TLCF}),(\ref{tr}) to the
description of spectral statistics at the Anderson transition.
The point
is that for RMEs considered here all states are critical.
For disordered electron systems there is a  mobility edge $E^{*}$,
and the states are nearly critical only in its vicinity where
$\xi(E)\propto |(E-E^{*})/E^{*}|^{-\nu}>L$.
Therefore the RMEs considered here correspond
formally to $\xi(E)=\infty$ and thus $\nu=\infty$. 
Indeed, 
it has been shown $^{8}$ that for finite $\nu$ the critical TLCF 
should have a
power-law tail
$R(s)=-c_{d}|s|^{-(1+\frac{1}{\nu d})}$, where $c_{d}\sim 
1/(\pi^2 \nu d)$. It vanishes in the limit $\nu\rightarrow \infty$
in agreement with the exponential decay of TCLF given by Eq.(\ref{TLCF}).
However, even for the realistic case $\nu\sim 1$ the power-law tail is
small due to the additional factor $\pi^2 d$. 

In conclusion, we link together three
different
random matrix ensembles suggested recently. Since in one of them the
eigenfunction statistics is known to be multifractal, we argue that
all three RME belong to the same universality class with the multifractal
eigenfunction statistics. 
By combining known solutions for all three ensembles we
suggest the form of the two-level correlation function in the region of
weak and extremely strong multifractality.   

{\bf Acknowledgments} We are grateful to B.L.Altshuler, C.M.Canali,
Y.V.Fyodorov, I.V.Lerner, A.D.Mirlin and
I.V.Yurkevich for numerous fruitful discussions. The hospitality 
extended to us at the University of Florida (V.E.K.) and ICTP,Trieste
(K.A.M.) is kindly 
acknowledged.
Work was supported in part by RFBR-INTAS grant No. 95-675, CRDF grant
No. RP1-209, and the Institute for Fundamental Theory at UF.


\begin{references}
\bibitem{23}
Unfolding is done by choosing  a new variable $\varepsilon=\varepsilon(E)$
from the relation $d\varepsilon/dE=\bar{\rho}(E)$.
\bibitem{27}
At $T/M=h\gg 1$, the grand-canonical ensemble used in Ref.[17] fails to
describe
correlations in a system of $M\rightarrow\infty$ free fermions at a
temperature $T\rightarrow\infty$, thus leaving open the problem of TLCF 
in RME, Eq.(\ref{III}), averaged over $\Omega$.
\end{references}
\end{document}